\documentclass[aps,showpacs,showkeys,preprint]{revtex4}

\begin{document}

\title{Matrix Algebras in Non-Hermitian Quantum Mechanics}

\author{Alessandro Sergi~\footnote{E-mail: sergi@ukzn.ac.zai
}
}

\affiliation{
School of Physics, University of KwaZulu-Natal, Pietermaritzburg Campus,
Private Bag X01 Scottsville, 3209 Pietermaritzburg, South Africa }

\begin{abstract}
In principle,
non-Hermitian quantum equations of motion can be formulated using
as a starting point either the Heisenberg's 
or the Schr\"odinger's picture of quantum dynamics.
Here it is shown in both cases 
how to map the algebra of commutators, defining the
time evolution in terms of a non-Hermitian Hamiltonian,
onto a non-Hamiltonian algebra with a Hermitian Hamiltonian.
The logic behind such a derivation is reversible, 
so that any Hermitian Hamiltonian
can be used in the formulation of non-Hermitian dynamics
through a suitable algebra of generalized (non-Hamiltonian) commutators.
These results provide a general structure (a template)
for non-Hermitian equations of motion to be used 
in the computer simulation of open quantum systems dynamics.
\end{abstract}



\maketitle

\today

It can be said that
the recent experimental realizations of ${\cal P T}$-symmetric 
time evolution by means 
of optical waveguides~\cite{optics,optics2} 
has renewed the interest in non-Hermitian quantum mechanics~\cite{nhqm}.
Besides practical applications 
to such recent developments~\cite{optics,optics2},
non-Hermitian theories are interesting since they can be thought of
as tools for computational  modelling
of open quantum systems dynamics~\cite{bf}.
It is known, in fact, that open quantum system dynamics can
be represented in terms of non-Hermitian quantum trajectory segments
interspersed by quantum transitions~\cite{bf}.
It is also worth mentioning that an analysis of the classical limit
of non-Hermitian quantum dynamics~\cite{cl} has revealed 
geometrical structures of phase space that are
represented in terms of dissipative brackets~\cite{grmela}.
Such brackets were originally proposed by Grmela~\cite{grmela}
and later widely discussed by Ottinger,
in his approach to beyond equilibrium thermodynamics~\cite{ottinger}.
Moreover, Ottinger himself has very recently provided an approach
to the quantization of such a dissipative dynamics~\cite{ottinger-qm}
by means of non-Hermitian quantum trajectories and jumps~\cite{bf},
hence providing some kind of logical \emph{short-circuit} to the 
quantum trajectory
representation of open quantum system dynamics~\cite{bf}.
Despite such interesting issues, so far attempts to characterize the general
features of non-Hermitian quantum dynamics have been few in number. 
A very interesting exception is the recent work of Graefe 
and Schubert~\cite{schubert}, where preliminary steps of 
such an analysis are performed
within a coherent-state representation.

In this brief communication, the focus is not on specific applications
but on the derivation of a general structure (a template) for
non-Hermitian equations of motion.
In full generality, quantum commutators defined in terms of a non-Hermitian
Hamiltonian are mapped
onto non-Hamiltonian commutators~\cite{sergi1,sergi2,sergi3}
defined in terms of the Hermitian part of the Hamiltonian.
Since commutators with the Hamiltonian naturally defines dynamics
in the Heisenberg picture, such a mapping allows one
to immediately identify very general characteristics
of non-Hermitian evolution in time.
This provides a conceptual connection between non-Hermitian
and non-Hamiltonian (or almost-Lie) dynamics.
In particular, the logic behind the mapping is reversible
so that any Hermitian Hamiltonian can be used in the formulation
of non-Hermitian dynamics by means of a suitable algebra
of generalized (non-Hamiltonian) commutators.
It is not difficult to foresee that this can
be exploited for devising novel algorithms
to simulate open quantum systems.
In fact, it is known that open quantum systems can be studied
by evolving the conserved Hamiltonian of the total system~\cite{attal}
and then measuring properties of the relevant subsystem only.
Analogously,
one would have to follow the non-Hermitian
evolution of the conserved Hamiltonian of the total system
and then calculate properties of just a relevant part.
The use of non-Hermitian evolution (in place of the natural
Hamiltonian dynamics) can provide the computational advantage
of reducing the number of degrees of freedom constituting
the bath of the relevant quantum subsystem~\cite{konstya}.
Such a logic is very well-known in the field
of classical molecular dynamics where one
can use the non-Hamiltonian evolution of a total conserved Hamiltonian
in order to impose thermodynamic constraints onto a relavant
subsystem~\cite{s1,s2,s3}.


Let us consider a non-Hermitian Hamiltonian operator
$ \hat{H} \ne \hat{H}^{\dag} $
determining the time evolution of an arbitrary quantum observable 
$\hat{\chi}$ in terms of the commutator
\begin{eqnarray}
i\hbar\frac{ d}{dt}\hat{\chi}
&=&[\hat{\chi},\hat{H}] 
=
\left[\begin{array}{cc} \hat{\chi} & \hat{H}\end{array}\right]
\cdot
\left[\begin{array}{cc} 0 & 1 \\ -1 & 0\end{array}\right]
\cdot
\left[\begin{array}{c} \hat{\chi} \\ \hat{H}\end{array}\right]
\;.\label{eq:comm}
\end{eqnarray}
Equation~(\ref{eq:comm}) displays the matrix structure of the
commutator, which was exploited in~\cite{sergi1}.
For future convenience, a more compact notation can be introduced.
To this end, defining the symplectic matrix
\begin{equation}
\mbox{\boldmath$\Omega$}=
\left[\begin{array}{cc} 0 & 1 \\ -1 & 0\end{array}\right]
\;,
\end{equation}
and the column vector
\begin{equation}
\mbox{\boldmath$\chi$}_{\hat{H}}
=\left[\begin{array}{c}\hat{\chi} \\ \hat{H}\end{array}\right]\;,
\label{eq:vecChiH}
\end{equation}
Eq.~(\ref{eq:comm}) can be written as
\begin{eqnarray}
i\hbar\frac{ d}{dt}\hat{\chi}
&= &\mbox{\boldmath$\chi$}_{\hat{H}}^{\rm T}\cdot
\mbox{\boldmath$\Omega$}\cdot
 \mbox{\boldmath$\chi$}_{\hat{H}}
\label{eq:matrix-comm} \\
&=&\left[\hat{\chi}(t), \hat{H}\right]_{\mbox{\small \boldmath$\Omega$}}
\label{eq:Lie-comm}
\;.
\end{eqnarray}
Equation~(\ref{eq:matrix-comm}) simply makes apparent 
the antisymmetric matrix structure of quantum evolution.
Such a structure underlies various non-Hamiltonian
and non-linear approximations 
of quantum dynamics~\cite{sergi1,sergi2,sergi3}.
Equation~(\ref{eq:Lie-comm}) simply shows that
the matrix structure is equivalent to the commutator,
defining a Lie bracket and a Lie algebra of operators in
quantum mechanics. The commutator is written putting 
the matrix $\mbox{\boldmath$\Omega$}$ into evidence.
This brings forward the idea~\cite{sergi1,sergi2,sergi3}
that more general algebras and brackets can be defined
simply by tampering with the definition of $\mbox{\boldmath$\Omega$}$.

The generic non-Hermitian Hamiltonian can always be split
in terms of an Hermitian ($\hat{H}_+$) and an anti-Hermitian part
($\hat{H}_-$):
\begin{eqnarray}
\hat{H}&=&\hat{H}_+ + \hat{H}_-\;,
\end{eqnarray}
where
$ \hat{H}_+=(1/2)\left(\hat{H} + \hat{H}^{\dag}\right)$
and $ \hat{H}_-=(1/2)\left(\hat{H} - \hat{H}^{\dag}\right)$.
In terms of such a decomposition,
the commutator with the Hamiltonian also splits into two parts:
$ [\hat{\chi},\hat{H}]= [\hat{\chi},\hat{H}_+]+ [\hat{\chi},\hat{H}_-] $.
It is now a matter of simple algebra to show that the 
commutator of an operator with the non-Hermitian Hamiltonian
is equivalent to a non-Hamiltonian commutator of the same
observable with only the Hermitian part of the Hamiltonian.
To this end, defining the antisymmetric matrix operator
\begin{equation}
\mbox{\boldmath$\Omega$}_{-+}
\equiv
\left[\begin{array}{cc} 0 & 1 +\hat{H}_-\left(\hat{H}_+\right)^{-1} \\
-1 - \left(\hat{H}_+\right)^{-1}\hat{H}_- & 0 \end{array}\right]
\;,
\end{equation}
one finds
\begin{eqnarray}
\left[\hat{\chi},\hat{H}\right]_{\mbox{\small\boldmath$\Omega$}}&\equiv& 
\mbox{\boldmath$\chi$}_{\hat{H}_+}^{\rm T}
\cdot
\mbox{\boldmath$\Omega$}_{-+}
\cdot
\mbox{\boldmath$\chi$}_{\hat{H}_+}\label{eq:nher-comm}
\\
&=&\left[\hat{\chi}(t),\hat{H}_+
\right]_{\mbox{\small \boldmath$\Omega$}_{-+}}
\label{eq:comm-Omega-+}
\;,
\end{eqnarray}
where in Eq.~(\ref{eq:nher-comm}), following an analogy 
with Eq.~(\ref{eq:vecChiH}),
the column vector
\begin{equation}
\mbox{\boldmath$\chi$}_{\hat{H}_+}
=
\left[\begin{array}{c} \hat{\chi} \\ \hat{H}_+\end{array}\right]
\label{eq:vecChiH+}
\end{equation}
has been defined.
Equation~(\ref{eq:comm-Omega-+}) introduces a
more general (non-Hamiltonian) commutator than the usual one based
on the symplectic matrix.
Equations~(\ref{eq:nher-comm})
and~(\ref{eq:comm-Omega-+}) realize a mapping between 
a standard commutator
in terms of a non-Hermitian Hamiltonian $\hat{H}$
and a non-Hamiltonian commutator defined in terms of $\hat{H}_+$, 
the Hermitian part of $\hat{H}$.

At this stage, it is worth discussing the starting point
of the derivation leading to the non-Hermitian matrix algebra
defined by Eqs.~(\ref{eq:nher-comm}) and~(\ref{eq:comm-Omega-+}).
Such a starting point is given by Eq.~(\ref{eq:comm}). 
This amounts to give a more fundamental role to the Heisenberg
law of evolution for observables, so that 
Eqs.~(\ref{eq:nher-comm}) and~(\ref{eq:comm-Omega-+}) 
defines an Heisenberg-based non-Hermitian  quantum dynamics.
Such a choice has some advantages. One of these is that
the Dirac's quantum-classical correspondence between commutators
and Poisson brackets remains unaltered~\cite{dirac}.

It must be remarked that, in order to formulate non-Hermitian
dynamics,
other authors have decided to give a more fundamental role
to the Schr\"odinger's picture, for example see~\cite{schubert},
and start from non-Hermitian equations of motion for state vectors:
\begin{equation}
\left\{\begin{array}{ccc}
|\dot{\Psi}\rangle &=& -\frac{i}{\hbar}\hat{H}_+|\Psi\rangle
+\frac{\hat{\Gamma}}{\hbar}|\Psi\rangle \\
\langle\dot{\Psi}| &=& \frac{i}{\hbar}\langle\Psi|\hat{H}_+
+\langle\Psi|\frac{\hat{\Gamma}}{\hbar}
\end{array}\right.
\;,
\end{equation}
where the Hermitian operator $\hat{\Gamma}=-i\hat{H}_-$ has been introduced.
These equations 
lead to an equation of motion for the density matrix
where an anticommutator, $[\ldots,\ldots]_+$, appears:
\begin{equation}
\frac{d}{dt}\hat{\rho}(t)
=-\frac{i}{\hbar}\left[\hat{H}_+,\hat{\rho}(t)\right]
+\frac{1}{\hbar}\left[\hat{\Gamma},\hat{\rho}(t)\right]_+
\;. \label{eq:schubert-rho}
\end{equation}
One can note that the equivalence between the Heisenberg and the
Schr\"odinger picture of dynamical evolution is lost
when the Hamiltonian is non Hermitian.
At this point, it should not be surprising that 
Eq.~(\ref{eq:schubert-rho}) can also be put in matrix form.
To this end, we can  define the general matrix operator 
(which is neither Hermitian nor antisymmetric)
\begin{equation}
\mbox{\boldmath$\Lambda$}
=\left[\begin{array}{cc}
0 & 1+i\hat{\Gamma}(\hat{H}_+)^{-1} \\
-1 + i (\hat{H}_+)^{-1}\hat{\Gamma} & 0
\end{array}\right]
\;.\label{eq:Lambda}
\end{equation}
Hence, Equation~(\ref{eq:schubert-rho}) can be rewritten as
\begin{eqnarray}
-i\hbar\frac{d}{dt}\hat{\rho}(t)
&=&\mbox{\boldmath$\rho$}_{\hat{H}_+}^T \cdot
\mbox{\boldmath$\Lambda$} \cdot
\mbox{\boldmath$\rho$}_{\hat{H}_+}\label{eq:Lambda-mo}
\\
&=& \left[\hat{\rho}(t),\hat{H}_+\right]_{\mbox{\small \boldmath$\Lambda$}}
\label{eq:Lambda-comm}
\;.
\end{eqnarray}
Equation~(\ref{eq:Lambda-comm})
introduces a bracket that has an antisymmetric
and a symmetric part and, hence, seems to directly
implement the original ideas of Grmela~\cite{grmela} 
within a quantum framework.
More in general, Equations~(\ref{eq:schubert-rho}), 
(\ref{eq:Lambda-mo}), and~(\ref{eq:Lambda-comm})
make an immediate contact with theories 
of dissipation~(\cite{grmela,ottinger})
and lead to interesting structures of phase space
in the classical limit~\cite{cl,schubert}.


What is interesting for open quantum system dynamics is that
the logic leading to Eq.~(\ref{eq:nher-comm}) can be reversed.
Consider an Hermitian Hamiltonian operator ${\cal H}={\cal H}^{\dag}$,
having by definition a spectrum of real eigenvalues.
Consider also a non-Hermitian operator $\hat{\xi}\neq\hat{\xi}^{\dag}$.
Upon defining an antisymmetric (but non-Hermitian) matrix 
$\mbox{\boldmath$\Omega$}_{\xi}$ as
\begin{equation}
\mbox{\boldmath$\Omega$}_{\xi}
=\left[\begin{array}{cc} 0 & \hat{\xi} \\ -\hat{\xi}^{\rm T} & 0
\end{array}\right]\;,
\end{equation}
one can define non-Hermitian evolution for any observable
in terms of a non-Hamiltonian commutator:
\begin{eqnarray}
i\hbar\frac{d}{dt}\chi
&=&\mbox{\boldmath$\chi$}_{\hat{\cal H}}^T
\cdot \mbox{\boldmath$\Omega$}_{\xi}\cdot
\mbox{\boldmath$\chi$}_{\hat{\cal H}}\label{eq:nham-nher}
\\
&=&\left[\hat{\chi}(t),\hat{\cal H}
\right]_{\mbox{\small\boldmath$\Omega$}_{\xi}}
\label{eq:comm-Omega-xi}
\;,
\end{eqnarray}
where the two-dimensional columns vector
\begin{equation}
\mbox{\boldmath$\chi$}_{\hat{\cal H}}
\equiv \left[\begin{array}{c} \hat{\chi} \\ \hat{\cal H}\end{array}\right]
\end{equation}
has been introduced in analogy with Eqs.~(\ref{eq:vecChiH}) and~(\ref{eq:vecChiH+}).
Equations~(\ref{eq:nham-nher}) and~(\ref{eq:comm-Omega-xi})
can be seen as a general way to define 
non-Hermitian dynamics for arbitrary Hamiltonian $\hat{\cal H}$
that (at time $t=0$) have a real spectrum of eigenvalues.
It can be argued that
the non-Hermitian time evolution introduced by the non-Hamiltonian
commutator in the right hand side of Eq.~(\ref{eq:nham-nher})
simulates the dynamics of an open quantum systems.
Such an inversion of logic can also be applied to 
the Schr\"odinger-based non-Hermitian dynamics.
Accordingly, Eqs.~(\ref{eq:Lambda-mo})
and~(\ref{eq:Lambda-comm}), together with the definition of the matrix
operator~(\ref{eq:Lambda}), can be reinterpreted as defining
a non-Hermitian algebra for an arbitrary Hamiltonian operator
with real eigenvalues at $t=0$.

In this brief communication, we have discussed the form
taken by the equations
of motion in non-Hermitian quantum mechanics
in both the Heisenberg's and the Schr\"odinger's picture.
In both cases, one can introduce a suitable algebra
of non-Hamiltonian commutators by exploiting 
the matrix structure of the equations of motion for quantum operators.
Inverting the logic, such algebras may be exploited for defining
the non-Hermitian evolution in terms of Hamiltonians
having a real spectrum of eigenvalues at the initial time.
The Heisenberg-based non-Hermitian dynamics preserves
Dirac's quantum-classical correspondence between commutators
and Poisson brackets~\cite{dirac}.
Moreover, already well-developed and tested numerical algorithms
for quantum dynamics~\cite{sergi-petruccio,sergi-ilya}
promise to be more readily applied
to such a formulation of non-Hermitian dynamics.
At the moment of writing, the classical limit of
the Heisenberg-based non-Hermitian quantum dynamics
remains to be investigated.
In particular, it is to be assessed whether
the phase space structure, found by Graefe~\cite{cl} e 
Schubert~\cite{schubert} within the Schr\"odinger-based
dynamics, also emerge from Heisenberg-based non-Hermitian 
quantum mechanics.

\section*{Acknowledgments}

This work is based upon research supported by
the National Research Foundation of South Africa.

The author is grateful to Prof. Paolo V. Giaquinta
for discussions and encouragement.

The author is also in debt with Dr. Kostya Zloschastiev
for criticism, suggestions, and a careful reading of the manuscript.


\end{document}